\documentstyle[12pt,epsfig]{article}
\begin{document}
\font\ninerm = cmr9
\def\footnoterule{\kern-3pt \hrule width
\hsize \kern2.5pt} \pagestyle{empty}
\vskip 0.5 cm

\begin{center}
{\large\bf  Some encouraging  and
some cautionary remarks on Doubly Special Relativity in
Quantum Gravity\footnote{\uppercase{T}alk given
at the ``10th \uppercase{M}arcel \uppercase{G}rossmann
\uppercase{M}eeting on \uppercase{G}eneral \uppercase{R}elativity"
(\uppercase{QG}5 session, \uppercase{R}io
de \uppercase{J}aneiro, \uppercase{J}uly 20-26, 2003).}}
\end{center}
\vskip 0.5 cm

\begin{center}
{\bf Giovanni Amelino-Camelia}\\
{\it $^a$Dipart.~Fisica, Univ.~Roma ``La Sapienza'' and INFN Sez.~Roma1\\
P.le Moro 2, 00185 Roma, Italy}
\end{center}

\vspace{1cm}
\begin{center}
{\bf ABSTRACT}
\end{center}

{\leftskip=0.6in \rightskip=0.6in
The idea of a role for DSR (doubly-special relativity)
in quantum gravity finds some encouragement
in a few scenarios, but in order to explore some key conceptual
issues it is necessary to find
a well-understood toy-quantum-gravity model that is fully compatible with
the DSR principles.
Perhaps the most significant source of encouragement comes from the
recent proposal
of a path for the emergence of DSR in Loop Quantum Gravity,
which however relies on a few assumptions on the results of
some computations that we are still unable to perform.
Indications in favor of the possibility of
using some elements of $\kappa$-Poincar\'e Hopf algebras
(and of the related $\kappa$-Minkowski noncommutative spacetime)
for the construction of a DSR theory have been discussed extensively,
but a few stubborn open issues must still be resolved, especially
in the two-particle sector.
It has been recently observed that certain structures encountered
in a formulation of 2+1-dimensional classical-gravity models
would fit naturally in a DSR framework,
but some key elements of these 2+1-dimensional models, including the description
of observers, might be incompatible with the DSR principles.}

\section{Introduction}
In Doubly-Special Relativity\cite{gacdsr} (DSR)
one looks for a transition in the Relativity postulates. A transition
which should be largely analogous to the Galilei $\rightarrow$ Einstein
transition.
Just like it turned out to be necessary, in order to describe high-velocity
particles, to set aside Galilei Relativity (with its lack of any characteristic
invariant scale) and replace it with Special Relativity
(characterized by the invariant velocity scale $c$),
it is at least plausible that,
in order to describe ultra-high-energy particles, we might have to set aside
Special Relativity and replace it with a new relativity theory,
a DSR, with two characteristic invariant scales, a new
small-length/large-momentum scale in addition to the familiar
velocity scale.

A theory will be compatible with the DSR principles if there
is complete equivalence of inertial observers (Relativity Principle)
and the laws of transformation between inertial observers
are characterized by two scales, a high-velocity scale and
a high-energy/short-length scale.
There are of course many observer-independent scales in physics
(the Planck constant $\hbar$, the electron mass (rest energy) $m_e$, $\dots$)
but from the perspective of the Relativity Principle in our present
theories there is one observer-independent
scale which plays a very special role:
the speed-of-light scale $c$ appears in the
rules of transformation\footnote{The
Relativity Principle prescribes that the laws of physics are
the same for all inertial observers,
but of course the same process is in general characterized by different measured
quantities in two different frames. The quantitative differences
between the measurement results of the two observers are connected
through the laws of transformations between inertial observers.}
between inertial observers. In a DSR theory one would have also
a second observer-independent scale with this property,
a short length scale $\lambda$ (large momentum scale $1/\lambda$).

Since in DSR one is proposing to modify the high-energy sector,
it is safe to assume that the present operative characterization
of the velocity scale $c$ would be preserved: $c$ is and should remain
the speed of massless low-energy particles.
Only experimental data could guide us toward the operative
description of the second invariant scale $\lambda$,
although its size is naturally guessed to be somewhere in the
neighborhood of the Planck length.
And, of course, it is very difficult to test the DSR idea in very general
terms. There are a few predictions that can be obtained using only the
general structure of the DSR Relativity principles.
For example, whereas in schemes for Planck-scale-{\underline{broken}}
Lorentz symmetry one naturally finds\cite{jaco,seth,dsrphen}
the characteristic prediction
that photons become unstable at sufficiently high energies,
in a DSR framework one may find that the second observer-independent
scale $\lambda$ affects the lifetimes of some particles, but necessarily
one must\footnote{If in a given theory, say, a photon is stable at low energies
but becomes unstable when its energy is higher than a certain value $E_*$
then clearly the theory is being formulated for a specific class of observers.
Two observers typically disagree on the energy of a particle but should not
disagree on whether or not a particle decays.}
have in DSR that a particle which is stable at low energies
is also stable at any however high energy.
However, with the exception of a few universal predictions (such as this one
pertaining to particle stability/instability),
different realizations of the DSR principles
(in particular, different postulates attributing operative meaning
to the second observer-independent scale $\lambda$)
will in general lead to different physical processes.
So one must in principle consider all possible formalisms
and physical pictures that could
provide a realization of the DSR idea.

Primarily inspired by a certain
analysis of some preliminary astrophysics
data\cite{agasa,aus,gactp}, the most
studied\cite{gacdsr,dsrnext,leedsr,gacdsrREV,dsrothers}
possibility  assumes that the second invariant $\lambda$
should acquire operative meaning by appearing in the dispersion relation
(the relation between energy and momentum of a particle of mass $m$),
which could take the form $E = f(p,m;c,\lambda)$.
The to-be-determined function $f$ must of course be such that for
small momenta $f \simeq \sqrt{c^2 p^2 + c^4 m^2}$.

In this specific type of DSR framework, which involves a $\lambda$-modified
dispersion relation, the ``discovery of DSR" would probably emerge
following a path which is analogous to the one of the discovery
of Special Relativity. The Michelson-Morley experiments can be viewed
as experiments providing evidence in favor of a $c$-modification
of the Galilei dispersion relation $E=p^2/2m$. At first the $c$-modification
was interpreted as a violation of the Relativity Principle ({\it i.e.}
a manifestation of the existence of a preferred class of inertial observers,
an ether), but then it was established that the $c$-modification is
observer independent: the Lorentz boost transformations are a modification
of the Galilei boost transformations such that
the dispersion relation $E = \sqrt{c^2 p^2 + c^4 m^2}$
holds in every inertial frame.
If evidence in favor of a dispersion relation of type  $E = f(p,m;c,\lambda)$
does emerge it would be natural to guess that $\lambda$ characterizes a
preferred class of inertial observers, but one should also consider
the possibility\cite{gacdsr} that instead the boost transformations
might be $\lambda$-modified without affecting in any way the Relativity Principle.
Of course, if the boost transformations are modified in order to render
invariant a dispersion relation of the type $E = f(p,m;c,\lambda)$
then one has repercussions also elsewhere in the relativistic theory,
and in particular the laws for the conservation of energy and momentum
must be also modified\cite{gacdsr,gacdsrREV} (in order to preserve their
status as covariant laws under boosts).

From a quantum-gravity perspective these DSR scenarios with modified
dispersion relation can be rather attractive, especially
because of the possibility to introduce a maximum momentum $1/\lambda$
(for example\cite{gacdsr,dsrnext,gacdsrREV} through a dispersion
relation such that $p \rightarrow 1/\lambda$ as $E \rightarrow \infty$).
And several attempts to introduce/find DSR in various descriptions of
quantum gravity and/or quantum spacetime have been made over these
past 3 years. Here I comment on some
of these studies and describe some possible general lessons that can
be drawn from them.

\section{The illustrative example of $\kappa$-Minkowski noncommutative spacetime}
Already the first studies of the DSR idea\cite{gacdsr,dsrnext}
explored the possibility that in the $\kappa$-Minkowski\cite{majrue,lukieAnnPhy}
noncommutative spacetime, characterized by coordinate
commutation relations $[x_j,t]= i x_j/\kappa~,~~[x_j,x_k]=0$,
one might be able to construct theories compatible with
the DSR principles.
Some results obtained in the analysis of $\kappa$-Minkowski do encourage the
possibility that the laws of  transformation between inertial observers
should be of a DSR-compatible type, with $\lambda \equiv 1/\kappa$.
It had been known for some time that $\kappa$-Minkowski
is dual (in an appropriate algebra sense) to
the much-studied $\kappa$-Poincar\'e\cite{lukieAnnPhy,lukie91e92} Hopf algebras.
Even though there are still some grey areas\cite{alessfranc}
in the physical characterization of a Hopf-algebra description
of spacetime symmetries,
this duality and the structure of the relevant Hopf algebra
can provide encouragement for
a modification of boosts that is admissible in DSR.

While the preceding $\kappa$-Poincar\'e  Hopf-algebra
literature had warned\footnote{Ref.~\cite{kpoinnogroup} observed
that by exponentiating $\kappa$-Poincar\'e generators one would
in general not obtain a symmetry group (only a ``quasigroup" in the
sense of Batalin\cite{batalin}).}
against\cite{kpoinnogroup}
attempts to integrate the boost generators to obtain a candidate for
finite boost transformations, the attempts\cite{gacdsr,dsrnext} to obtain
a DSR physical theory on the basis of $\kappa$-Poincar\'e mathematics
could not avoid a description of finite boosts, and in the investigation
of these issues it emerged that some descriptions of the $\kappa$-Poincar\'e
Hopf algebra do allow (by exponentiation of the $\kappa$-Poincar\'e
generators) a perfectly consistent implementation of finite boosts
on one-particle systems.
And these recent investigations also showed that
it is rather natural to introduce a maximum value for momentum (and/or
energy), {\it i.e.} to obtain a description of boosts that saturates to
the maximum value $p_{max} = \kappa = 1/\lambda$.
The picture of energy-momentum space admits description
in terms of a deSitter-type geometry\cite{jurekDS}, at least in the sense that
one can find a straightforward map between the energy-momentum variables
and the coordinates of a deSitter-type spacetime.

Because of the mentioned duality,
these features of $\kappa$-Poincar\'e should be applicable to
theories in $\kappa$-Minkowski spacetime,
and this is of encouragement for a DSR formulation of
physics in $\kappa$-Minkowski.
But the duality with $\kappa$-Poincar\'e is also the source of some concern
for a DSR description of multi-particle systems in $\kappa$-Minkowski.
In fact, while in the one-particle sector everything can be formulated
in a DSR-compatible way ({\it i.e.} one can find $\kappa$-Poincar\'e mathematical
structures which can be used to construct a DSR physical theory),
already for a simple system of two particles there is no known way
to use $\kappa$-Poincar\'e mathematics to obtain a DSR-compatible formulation.
In particular, the law of energy-momentum conservation which is
advocated by $\kappa$-Poincar\'e experts\cite{lukieNEWdsr} is incompatible
with the DSR principles, since it combines with the $\kappa$-Poincar\'e
dispersion relation in such a way to lead to the emergence of
a preferred class of inertial observers\cite{gac3perspeAreanew,dsrphen}.
Perhaps the solution of this problem will simply require us to uncover some
new structures within the $\kappa$-Poincar\'e Hopf algebra
mathematics, but at present this can only be conjectured.

\section{Proceeding with caution}
 $\kappa$-Minkowski has been the first example of quantum spacetime
considered from a DSR perspective. Although, as just stressed,
it is still unclear whether $\kappa$-Minkowski really is fully compatible
with the DSR principles, it is a good test case to explain how
DSR-type structures might emerge in a quantum spacetime.
One way to see this is through the analysis of plane waves
in $\kappa$-Minkowski, $e^{i k x} e^{-i k_0 t}$,
where it is clear that the composition of two waves,
$e^{i q x} e^{-i q_0 t} = e^{i k x} e^{-i k_0 t} e^{i p x} e^{-i p_0 t}$,
cannot involve a linear combination of momenta ($q=k+p$ is incompatible
with the relations $[x_j,t]= i x_j/\kappa~,~~~[x_j,x_k]=0$).
If $\kappa$-Minkowski is eventually understood as a DSR-compatible spacetime
it is likely that this nonlinear law of composition of wave exponentials
will play a key role in the analysis.

A few other examples of quantum gravity or quantum-spacetime
pictures which invite one to consider the DSR possibility have been found.
But at present the key challenge for this research programme
is to show that at least in one specific context DSR is
actually present, a specific example of a framework that is fully compatible
with the DSR principles.
As presently understood $\kappa$-Minkowski, because of the key issues
that remain unsettled, still cannot be viewed as such a fully-worked-out
DSR example.
And among the other ``DSR candidates" the situation is similar:
some features that provide support for the idea of a DSR description
have been uncovered, but the full compatibility with the DSR
principles has not been established.
We are however learning more about what is needed and what is not needed for
(or is incompatible with) a DSR framework, and in setting up future research
on this subject these lessons might be precious.

\subsection{DSR in classical spacetime not likely}
If indeed one is looking for the specific type of DSR framework
which involves nonlinearities in the energy-momentum sector
it seems unlikely that one should be able to construct the theory
in a classical spacetime ({\it i.e.} a spacetime with sharp
localization of events). In a flat classical spacetime plane waves will combine
in the familiar straightforward way (unlike the case of $\kappa$-Minkowski
noncommutativity, which, as mentioned, inevitably leads to
nonlinearities of the composition of waves). In such contexts
nonlinearities in the transformation laws of energy and momentum
could be introduced only in a rather fictitious way ({\it e.g.} by
an otherwise unjustified redefinition of the energy and momentum variables
starting from the standard special-relativistic case), and, chances are,
at some point the formalism will remove the nonlinearities at the level
of truly observable predictions.

Even if one attempts to introduce the two scales directly in spacetime
structure (rather than starting from an energy-momentum space intuition)
some difficulties should emerge. Some authors have placed much
emphasis of the fact that Fock, long ago\cite{fock},
stumbled upon the observation that in a flat spacetime one obtains
transformation laws that involve two scales upon renouncing to
the objectivity of parallelism of worldlines.
Fock was actually thinking in reverse: in an analysis of the conceptual
structure of Special Relativity he explored the implication
of the removal of one of the hypothesis of Special Relativity,
the one that establishes that two worldlines are parallel for all observers
if they are parallel for one observer,
and found that this would allow a two-scale family of transformation laws.
But, while conceptually insightful, this result did not gain
any interest from a physics perspective (not even in the eyes
of Fock himself\cite{fock}) since, as one should expect, the second scale
ended up being necessarily a large length scale (removed in the
large-distance limit rather than the short-distance limit),
a possibility which can be safely excluded on the basis of our abundant
low-energy data, which are all fully compatible with ordinary (one-invariant-scale)
Special Relativity.
It seems that essentially Fock might have rediscovered deSitter spacetime:
his work started from the assumption of a flat spacetime but by
removing the objective parallelism of worldlines, and assuming that
spacetime could be described in terms of a classical geometry, he effectively
introduced constant curvature in spacetime.

\subsection{Hopf-algebra description of Poincar\'e symmetries
not sufficient for DSR}
As stressed in the previous section, when I considered the possibility of
obtaining a DSR theory in $\kappa$-Minkowski spacetime
using $\kappa$-Poincar\'e mathematics,
a two-scale Hopf-algebra generalization of the Poincar\'e Lie algebra
does not automatically provide the mathematical ingredients to
formulate a DSR theory.
Perhaps as a result of the focus on mathematics that permeates all of
quantum-gravity research this point is often missed.
An algebra involves a certain collection of mathematical
structures which admit in principle a large number of possible uses
in the construction of physical theories.
The example of the $\kappa$-Poincar\'e Hopf algebra should help clarify
this point: there are plenty of structures in
the $\kappa$-Poincar\'e/$\kappa$-Minkowski
literature ({\it e.g.} the generators, the coproducts, the antipodes...)
but in physics one must find a way to use these structures in the description
of various physical features of the relativistic theory.

In the simplest cases the mathematics just confronts us with some
alternatives, and the physicists select the structures on the basis
of compatibility with certain physics principles. For example in the
description of fields in $\kappa$-Minkowski one must of course
introduce a differential calculus, for which various alternatives have
been proposed in the $\kappa$-Minkowski literature.
In particular there has been interest in a four-dimensional\cite{gacmajoeck}
and in a five-dimensional\cite{Sitarz} differential calculus,
and both of these differential calculi deserve equal consideration
at the mathematics level, but if one
insists on compatibility with the DSR principles only the
five-dimensional differential calculus turns out~\cite{alessmich}
to be acceptable.

And one cannot exclude that among the many alternatives provided
by the mathematics side there might not be a single one that accomplishes
the tasks required by the physicist.
For example
in the search of a DSR-compatible law of
energy-momentum conservation physicists should choose
among a large variety of different structures on the mathematics side which are
plausible candidates
as building blocks for the law of energy-momentum conservation. And, as mentioned,
it is still unclear whether any
of the mathematical structures which emerged in
the $\kappa$-Poincar\'e literature
would allow to introduce a law of conservation
of energy-momentum that is compatible with the DSR principles (at least it is
at present not known that any of these structures would lead to a DSR-compatible
conservation law).

\subsection{deSitter geometry for energy-momentum space
not sufficient for DSR}
Just like a superficial look at $\kappa$-Minkowski might lead to the
incorrect expectation that the availability of a two-scale
Poincar\'e-like algebra would be automatically sufficient for
a DSR formulation, in turn the observation that $\kappa$-Poincar\'e,
in an appropriate sense, predicts an energy-momentum space which has
deSitter-type geometry might lead to the
incorrect expectation that whenever it is possible to find a natural-looking
map from the energy-momentum variables to some coordinates over a deSitter
geometry one should be able to find a DSR formulation of the theory.
Of course, this expectation is also incorrect. It is true that a natural
way to introduce a second relativistic scale, in the sense of DSR,
can be the one of an energy-momentum space with observer-independent
constant curvature, but the observer-independence of the
curvature is not assured by the existence of a map connecting the energy-momentum
variables (as measured by a given observer)
to some coordinates on a deSitter geometry.

In order to make this remark more concrete let me propose a simple
analogy. The propagation of light in a water-pool is (to very good approximation)
described by a dispersion relation $E = \sqrt{c_{water}^2 p^2 + c_{water}^4 m^2}$
(of course, $m=0$ for photons)
which of course allows a map from the energy-momentum variables to
some coordinates on a Minkowski geometry.
But we know that the scale $c_{water}$ is not observer independent.

From the DSR perspective the problem is even more serious: assuming
the space of one-particle energy-momentum is {\underline{truly}}
deSitter-like (with observer-independent curvature)
it remains to be seen how the space of multiparticle momenta
should be described. If indeed the DSR proposal involves nonlinearities
in energy-momentum space the step from the one-particle sector
to the multi-particle sector will inevitably involve some delicate issues.

The fact that the presence of a deSitter-type geometry for energy-momentum space
cannot be used to fully specify the symmetry properties was already rather
clear to Snyder in the 1940s. Snyder was looking\cite{snyder} for a
nonclassical description of spacetime which would regulate the UV divergences
of quantum field theory while being fully compatible with
ordinary\footnote{This renowned paper by Snyder\cite{snyder} is one of
the most cited in the
noncommutative-geometry literature, but apparently it is often cited
without reading, and this is leading to frequent misrepresentations of the
objectives and the results of the paper. For example, it is sometimes said that
Snyder proposed a modification of Lorentz symmetry, and this is
rather paradoxical since,
on the contrary, Ref.~\cite{snyder} provides a clear statement of
objectives, very early in the analysis, in which the preservation of ordinary
Lorentz symmetry is stressed as a key point.}
(undeformed) Lorentz symmetry, and Snyder used a
strategy which automatically implied
that energy-momentum space would have a deSitter-type geometry.

\section{Special caution in 2+1-dimensional spacetime}
Two recent papers\cite{kodadsr,jurekkodadsr}
have considered the possibility that the description of gravity
in 2+1-dimensional spacetime might provide a good toy model
in which to construct a DSR theory.
Before discussing these proposals
I find it necessary to stress that from a DSR
perspective the 2+1 context might not provide the correct intuition
for the 3+1 context.
In the study of most types of theories the 2+1 context contains the
same logical structure as the 3+1 context, but with a welcome
reduction in the level of technical complexity.
But from a DSR perspective the logical structure of theories in
2+1 dimensions may be significantly different from the one
of their 3+1-dimensional counterparts.
This can be seen already at the simple level of dimensional analysis
of the scales involved in the theory.
Whereas in 3+1 dimensions both the Planck length and the Planck energy
are related to the gravitational constant through the Planck constant
($L_p \equiv \sqrt{\hbar G/c^3}~,~~E_p \equiv \sqrt{\hbar c^5/ G}$),
in 2+1 dimensions the (2+1 version of the)
Planck energy is obtained only in terms of the speed-of-light scale
and the gravitational constant: $E_p^{(2+1)} \equiv c^4/ G^{(2+1)}$.
(The Planck length however is still introduced through the Planck
constant, $L_p^{(2+1)} \equiv \hbar c^3/G^{(2+1)}$.)
Since a possible novel operative meaning for the Planck energy and/or
length is the key issue under investigation in DSR this observation
could be important.

\subsection{Cautionary remarks on DSR from
2+1D quantum gravity with q-deSitter symmetry algebra}
At least in some formulations of quantum gravity in 2+1 spacetime
dimensions a q-deformed deSitter symmetry algebra $SO(3,1)_q$
emerges\cite{roche}
for nonvanishing cosmological constant, and the relation between the
cosmological constant $\Lambda$ and
the $q$ deformation parameter
takes the form $\ln q \sim \sqrt{\Lambda} L_p$
for small $\Lambda$. It was observed in Ref.~\cite{kodadsr}
(using a well-established result on the contractions of the q-deSitter
algebra\cite{lukie91e92})
that this relation $\ln q \sim \sqrt{\Lambda} L_p$
implies that the flat-spacetime limit ($\Lambda \rightarrow 0$)
is not described by a Poincar\'e Lie algebra but rather by
a $\kappa$-Poincar\'e Hopf algebra.
This provided the first ever argument in favor of the possibility
of the emergence of DSR in 2+1D quantum gravity. But the analysis
only shows that the $\kappa$-Poincar\'e Hopf algebra should have a role
in the flat-spacetime limit, without providing a fully physical picture
of this role. And in any case, as stressed above, the presence of
a $\kappa$-Poincar\'e Hopf algebra somewhere in the formalism of the theory,
while providing automatically some ingredients that are suitable
for DSR relativity, also introduces some unsolved issues for the consistency
with the DSR principles. So the analysis reported in Ref.~\cite{kodadsr}
rigorously shows that the flat-spacetime limit in certain formulations\cite{roche}
of 2+1D quantum gravity is not described by classical (Lie-algebra)
Poincar\'e symmetries, but, pending the mentioned open issues concerning
the possibility of using $\kappa$-Poincar\'e mathematics in the construction of DSR
physical theories, it is not really conclusive concerning the emergence
of a DSR framework.

\subsection{Cautionary remarks on DSR in the {\underline{classical}}
2+1D gravity of Matschull et al}
In the more recent Ref.~\cite{jurekkodadsr}
it was also observed that
some aspects of a certain formulation of classical gravity
for point particles in 2+1 dimensions, mostly due to
Matschull {\it et al}\cite{mats1,mats2,mats3},
are compatible with the DSR idea.
The key DSR-friendly ingredients are the presence of a maximum value
of mass and a description of energy-momentum space which admits
mapping onto the coordinates of a deSitter geometry.
These are rather striking observations, and Ref.~\cite{jurekkodadsr}
presents an elegant argument which rather compellingly raises the
possibility of a DSR formulation of the framework developed
by Matschull {\it et al}.
However, as for the proposal discussed in the previous subsection,
also in this case several additional results must be obtained
in order to verify whether or not such a DSR formulation is possible.
One key point is that Matschull {\it et al} formulate\cite{mats1,mats2}
the theory from the very
beginning by making explicit reference to a specific frame,
the frame of the center of mass of the multiparticle system.
There is therefore no natural reason to assume that the features that
emerge from the analysis are observer independent.
And, as stressed above, features like the presence of a maximum energy
and of a deSitter-geometry-compatible structure in energy-momentum space
can emerge in equally natural manner in a theory with a preferred frame
and in a DSR theory.

Moreover, rather than a deformation of the 6 translation/rotation/boost
classical (Lie-algebra) symmetries of 2+1D space, many aspects
of the theory, because of an underlying conical geometry, appear to be
characterized by only two symmetries: a rotation and a time translation.
The framework of Matschull {\it et al} has a first level of formulation
describing (topological) gravity interactions among particles
in a classical Minkowski 3D spacetime. But the level of formulation where
one is here attempting
to see the emergence of a DSR framework is the formulation
in which this interacting
system is turned into a a system of free particles in a gravity-modified
geometry (turning dynamics into kinematics by changing geometry).
And at that level indeed the
framework of Matschull {\it et al} must be described in terms
of conical geometry,
characterized only by a rotation and a time translation symmetry transformations.

The description in terms of conical geometry is also closely related
to the fact that the ``observers at infinity"
in the framework of Matschull {\it et al} do not really decouple from
the system under observation. These observers might not be good examples
for testing the Relativity Principle.

For the conjecture of emergence of a DSR framework it is
also puzzling that, especially when considering particle collisions,
Refs.~\cite{mats1,mats2,mats3} appear to describe
frequently as total momentum of a multiparticle system simply the sum
of the individual momenta of the particles composing the system.
From a DSR perspective,
such a linear-additivity law for total momentum is of course
incompatible\cite{gacdsr} with
deformed laws of transformation of energy-momentum.
So it appears that this theory of classical-gravity
interactions among point particles in 2+1 dimensions,
while providing important intuition for other aspects of quantum-gravity
research, is likely to fail to be useful in DSR research, unless we manage
to introduce some significantly new elements with respect to the
original formulation of Refs.~\cite{mats1,mats2,mats3}.

\section{A path for DSR in Loop Quantum Gravity}
While I am here placing strong emphasis on some open issues
that confront the further development of DSR research,
clearly the robust results we already have are very significant
and go well beyond what one could have expected\cite{gacdsr}
as the results of only 3 years of work.
And we now even see a truly remarkable opportunity for DSR:
as observed in
the later part of the analysis reported in Ref.~\cite{kodadsr}
there is a possible path for the emergence of
a role for DSR in Loop Quantum Gravity, which is one of the
most popular approaches to the quantum-gravity problem.

\subsection{The Kodama state, q-deSitter, and DSR}
The path for the emergence of
a role for DSR in Loop Quantum Gravity proposed
in Ref.~\cite{kodadsr} is closely analogous to the one described
here in Subsection~4.1 for one of the formulations
of 2+1D quantum gravity.
In fact, also in the 3+1D context of Loop Quantum Gravity
the literature presents some support for the presence of a
q-deformation of the deSitter symmetry algebra when
there is nonvanishing cosmological constant.
These arguments are based mainly on the properties of the Kodama state\cite{kodama}
and on some approaches to the formulation of boundary observables
in Loop Quantum Gravity\cite{artem}.
As discussed in Ref.~\cite{leekoda} (which had argued for
a mechanism rather similar to the one then developed in
Ref.~\cite{kodadsr}, but without the elements concerning the
symmetry-algebra analysis)
and Ref.~\cite{kodadsr},
in the 3+1D context one expects a renormalization of energy-momentum
which is still not under control in the Loop-Quantum-Gravity
literature.
For the analysis of Ref.~\cite{kodadsr}
this essentially turns into an inability to fully predict
the relation between the $q$-deformation parameter
and the cosmological constant,
but for small $\Lambda$
one should expect $\ln q \sim (\sqrt{\Lambda} L_p)^r$, with $r$
a numerical parameter to be determined through the mentioned renormalization
procedure.
If the choice $r=1$ turns out to be correct one would find
again (as for the formulation
of 2+1D quantum gravity mentioned in Subsection~4.1)
that the flat-spacetime limit, $\Lambda \rightarrow 0$,
is characterized by a $\kappa$-Poincar\'e Hopf algebra
(and in turn, assuming the issues mentioned in Section 2 find a
positive solution, this could lead to a DSR formulation).

\subsection{A new Planck-scale Cosmology even without DSR}
I also want to stress that
the hypothesis $\ln q \sim (\sqrt{\Lambda} L_p)^r$
for a $q$-deformed deSitter algebra in Loop Quantum Gravity
is actually interesting
even if the choice $r=1$ turns out not to be correct.
For $r>1$ the flat-spacetime limit would be characterized by a classical
Poincar\'e algebra, but the $q$-deformation would still be significant,
with potentially interesting consequences in phenomenology
and cosmology, whenever $\Lambda \neq 0$ (or there is a
nonvanishing curvature scalar).



\begin{thebibliography}{0}

\bibitem{gacdsr} G.~Amelino-Camelia, gr-qc/0012051,
{\it Int.~J.~Mod.~Phys.}~{\bf D11}, 35 (2002);
hep-th/0012238,
{\it Phys.~Lett.}~{\bf B510}, 255 (2001).

\bibitem{jaco} T.~Jacobson, S.~Liberati and D.~Mattingly,
hep-ph/0112207, {\it Phys.~Rev.}~{\bf D66}, 081302 (2002).

\bibitem{seth} T.J.~Konopka and S.A.~Major,
{\it New J.~Phys.}~{\bf 4}, 57 (2002).

\bibitem{dsrphen} G.~Amelino-Camelia, J.~Kowalski-Glikman,
G.~Mandanici and A.~Procaccini,
gr-qc/0312124.

\bibitem{agasa} M.~Takeda {\it et al},
{\it Phys.~Rev.~Lett.}~{\bf 81}, 1163 (1998);

\bibitem{aus} R.J.~Protheroe and H.~Meyer,
{\it Phys.~Lett.}~{\bf B493}, 1 (2000).

\bibitem{gactp} G.~Amelino-Camelia and T.~Piran,
astro-ph/0008107,
{\it Phys.~Rev.}~{\bf D64}, 036005 (2001);
G.~Amelino-Camelia,
gr-qc/0012049, {\it Nature} {\bf 408}, 661 (2000).

\bibitem{dsrnext} J.~Kowalski-Glikman,
hep-th/0102098,
{\it Phys.~Lett.}~{\bf A286}, 391 (2001);
R.~Bruno, G.~Amelino-Camelia and J.~Kowalski-Glikman,
hep-th/0107039,
{\it Phys.~Lett.}~{\bf B522}, 133 (2001);
G.~Amelino-Camelia, D.~Benedetti, F.~D'Andrea and
A.~Procaccini,
hep-th/0201245,
{\it Class.~Quant.~Grav.}~{\bf 20}, 5353 (2003);
J.~Kowalski-Glikman and S.~Nowak,
hep-th/0204245.

\bibitem{leedsr} J.~Magueijo and L.~Smolin,
hep-th/0112090, {\it Phys.~Rev.~Lett.}~{\bf 88}, 190403 (2002);
gr-qc/0207085, {\it Phys.~Rev.}~{\bf D67},  044017 (2003).

\bibitem{gacdsrREV} G.~Amelino-Camelia,
gr-qc/0207049, {\it Nature} {\bf 418}, 34 (2002);
gr-qc/0210063, {\it Int.~J.~Mod.~Phys.}~{\bf D11}, 1643 (2002).

\bibitem{dsrothers} S.~Mignemi,
hep-th/0208062;
M.~Toller,
hep-ph/0211094;
A.~Chakrabarti,
hep-th/0211214, {\it J.~Math.~Phys.}~{\bf 44}, 3800 (2003).
S.~Mignemi,
gr-qc/0304029, {\it Phys.~Rev.}~{\bf D68},  065029 (2003);
A.~Ballesteros, N.R.~Bruno and F.J.~Herranz,
hep-th/0305033, {\it J.~Phys.}~{\bf A36}, 10493 (2003).
G.~Svetlichny,
hep-th/0305100;
A.~Ballesteros, N.R.~Bruno and F.J.~Herranz,
hep-th/0306089, {\it Phys.~Lett.}~{\bf B574},  276 (2003);
S.K.~Kim, S.~M.~Kim, C.~Rim and J.H.~Yee,
gr-qc/0401078.

\bibitem{majrue} S.~Majid and H.~Ruegg,
{\it Phys.~Lett.}~{\bf B334}, 348 (1994).

\bibitem{lukieAnnPhy} J.~Lukierski, H.~Ruegg and W.J.~Zakrzewski
{\it Ann.~Phys.}~{\bf 243}, 90 (1995).

\bibitem{lukie91e92} J.~Lukierski, H.~Ruegg, A.~Nowicki
and V.N.~Tolstoi,
{\it Phys.~Lett.}~{\bf B264}, 331 (1991);
J.~Lukierski, A.~Nowicki and H.~Ruegg,
{\it Phys.~Lett.}~{\bf B293}, 344 (1992).

\bibitem{alessfranc} A.~Agostini, G.~Amelino-Camelia, F.~D'Andrea,
hep-th/0306013.

\bibitem{kpoinnogroup} J.~Lukierski, H.~Ruegg and W.~Ruhl,
{\it Phys.~Lett.}~{\bf B313}, 357 (1993).

\bibitem{batalin} I.A.~Batalin,
{\it J.~Math.~Phys.}~{\bf 22}, 1837 (1981).

\bibitem{jurekDS} J.~Kowalski-Glikman and S.~Nowak,
hep-th/0304101,
{\it Class.~Quant.~Grav.}~{\bf 20}, 4799 (2003).

\bibitem{lukieNEWdsr} J.~Lukierski and A.~Nowicki, hep-th/0203065.

\bibitem{gac3perspeAreanew} G.~Amelino-Camelia,
gr-qc/0205125; gr-qc/0309054.

\bibitem{fock} See the appendix of V.~Fock, ``The theory
of space-time and gravitation" (Pergamon Press, 1964).

\bibitem{gacmajoeck} S.~Majid and R.~Oeckl, math.QA/9811054;
G.~Amelino-Camelia and S.~Majid,
hep-th/9907110,
{\it Int.~J. Mod.~Phys.}~{\bf A15}, 4301 (2000).

\bibitem{Sitarz} A.~Sitarz,
hep-th/9409014,
{\it Phys.~Lett.}~{\bf B349},  42 (1995);
C.~Gonera, P.~Kosinski and P.~Maslanka,
q-alg/9602007.

\bibitem{alessmich} A.~Agostini, G.~Amelino-Camelia, M.~Arzano,
gr-qc/0207003.

\bibitem{snyder} H.S.~Snyder, {\it Phys.~Rev.}~{\bf 71}, 38 (1947).

\bibitem{kodadsr} G.~Amelino-Camelia, L.~Smolin and A.~Starodubtsev,
hep-th/0306134

\bibitem{jurekkodadsr} L.~Freidel, J.~Kowalski-Glikman and L.~Smolin,
hep-th/0307085.

\bibitem{roche} K.~Noui and P.~Roche,
gr-qc/0211109.

\bibitem{mats1} H.-J.~Matschull and M.~Welling,
gr-qc/9708054, {\it Class.~Quant.~Grav.}~{\bf 15}, 2981 (1998).

\bibitem{mats2}
J.~Louko and H.-J.~Matschull,
gr-qc/9908025, {\it Class.~Quant.~Grav.}~{\bf 17}, 1847 (2000).

\bibitem{mats3} H.-J.~Matschull,
gr-qc/0103084,
{\it Class.~Quant.~Grav.}~{\bf 18}, 3497 (2001);
J.~Louko and H.-J.~Matschull,
gr-qc/0103085,
{\it Class.Quant.Grav.}~{\bf 18}, 2731 (2001).

\bibitem{kodama} H.~Kodama,
{\it Phys.~Rev.}~{\bf D42}, 2548 (1990).

\bibitem{artem} A.~Starodubtsev,
hep-th/0306135.

\bibitem{leekoda} L.~Smolin,
hep-th/0209079.


\end{thebibliography}
\end{document}